\documentclass[8pt,a4paper,twocolumn]{article}

\usepackage[switch]{lineno} 
\usepackage{multicol,lipsum}

\usepackage[]{cite}
\usepackage{csquotes}
\usepackage{subfiles}

\usepackage{abstract}

\usepackage[table, dvipsnames]{xcolor}
\definecolor{lightgray}{gray}{0.85}

\usepackage[top=1.8cm,left=1.7cm,right=1.7cm,bottom=1.75cm]{geometry}
\usepackage{relsize}
\usepackage{tabularx}
\usepackage{array}
\usepackage{wrapfig}
\usepackage{multirow}
\usepackage{booktabs}
\usepackage{caption}
\usepackage{graphicx}
\usepackage{xcolor}
\usepackage{comment}

\usepackage{titlesec}
\titlespacing*{\section}{0pt}{0.1\baselineskip}{0.2\baselineskip}

\usepackage[affil-it]{authblk} 
\usepackage{etoolbox}

\usepackage{lmodern}
\usepackage{amsmath}
\usepackage{amsfonts}
\usepackage{mathtools}

\makeatletter
\patchcmd{\@maketitle}{\LARGE \@title}{\fontsize{16}{19.2}\selectfont\@title}{}{}
\makeatother

\begin{document}

\title{\textbf{{Birefringence-free photoelastic modulator with centimeter-square aperture operating at 2.7 MHz with sub-watt drive power}}}

\author[1*]{Okan Atalar}
\author[1]{Amin Arbabian}

\affil[1]{\textit{Department of Electrical Engineering, Stanford University, Stanford, California 94305, USA}}
\affil[*]{Corresponding author: okan@stanford.edu\vspace{-2em}}

\date{}
\twocolumn[
  \begin{@twocolumnfalse}
\maketitle

\thispagestyle{empty}

\begin{abstract}
Photoelastic modulators are optical devices with a broad range of applications. These devices typically utilize a transverse interaction mechanism between acoustic and optical waves, resulting in a fundamental trade-off between the input aperture and the modulation frequency. Commercially available modulators with centimeter square apertures have operating frequencies in the vicinity of 50~kHz. In this work, we experimentally demonstrate a birefringence-free photoelastic modulator operating at approximately 2.7~MHz with a centimeter square aperture, increasing the operating frequency substantially compared to existing approaches. Using the modulator and polarizers, we demonstrate close to $\pi$ radians polarization modulation amplitude with sub-watt drive power, translating to nearly 100\% intensity modulation efficiency at the fundamental (2.7~MHz) and second harmonic (5.4~MHz) frequencies.
\vspace{2em}
\end{abstract}

\end{@twocolumnfalse}
]

Photoelastic modulators are acousto-optic devices typically constructed by bonding an isotropic material to a piezoelectric transducer~\cite{1966_pem,improved_ellipsometry_pol_mod,piezo_optical_birefringence_mod}. Acoustic resonance with a transverse interaction mechanism is utilized to realize low-power, wide-angle, and birefringence-free polarization modulation. They are used for applications including polarimetry, ellipsometry, polarization spectroscopy, linear and circular dichroism, and intensity modulation of free-space beams~\cite{piezo_optical_birefringence_mod,mueller_matrix_polarimetry,pem_for_polarimetry_ellipsometry,improved_ellipsometry_pol_mod,applications_pem,stokes_and_mueller_polarimetry,rapid_wide_field_Mueller,elimination_artifact_circular_dichroism,rapid_time_gated_polarimetric_Stokes,spectropolarimetric,45_degrees_PEM_drive}.  
The transverse interaction mechanism results in a fundamental trade-off between the input aperture and the modulation frequency. Given the commonly used materials for the isotropic interaction medium, such as silica, zinc selenide, and calcium fluoride~\cite{pem_infrared,pem_infrared_2,Hinds_Instruments}, the modulation frequencies are typically in the vicinity of 50~kHz for centimeter square apertures.

We have recently demonstrated a new class of optical modulators referred to as \textquote{longitudinal piezoelectric resonant photoelastic modulators}~\cite{longitudinal_nat_paper,yz_ln_paper,optically_isotropic_atalar}. These devices consist of a single crystal functioning as both the piezoelectric transducer and the acousto-optic interaction medium. They employ a collinear acousto-optic interaction mechanism, overcoming the trade-off between input aperture and modulation frequency. This allows simultaneous access to high modulation frequencies and input apertures. By choosing a crystal that is optically isotropic and piezoelectric (non-centrosymmetric cubic) to construct the modulator, birefringence-free photoelastic modulators were demonstrated in proof-of-concept experiments exhibiting a wide acceptance angle~\cite{optically_isotropic_atalar}.

In this work, using similar design principles from our previous proof-of-concept demonstration, we demonstrate a birefringence-free photoelastic modulator operating at approximately 2.7~MHz with centimeter-square aperture and with sub-millimeter thickness. Owing to the high quality factor ($Q$) and mode purity of the excited acoustic standing wave in the modulator, we demonstrate substantial polarization modulation for 940~nm wavelength light with sub-watt drive power. The modulation efficiency of the modulator reported in this work is more than an order of magnitude higher compared to our previous demonstration.

We fabricate the modulator using an undoped, double-side polished GaAs wafer of thickness 522~$\mu $m, 50.8~mm diameter, and with crystal orientation (332) in Miller indices notation. We purchased the wafers from DOWA Electronics Materials Co., Ltd., a company that specializes in GaAs wafer production. The purity of the crystal is important to achieve high $Q$ and thus high modulation efficiencies for sub-watt drive powers. Low-power operation is critical for an optical modulator to avoid bulky radio frequency (RF) drivers and to be practical for time-of-flight applications~\cite{ToF_atalar}.

\begin{figure}[ht]
\centering
\includegraphics[width=\linewidth]{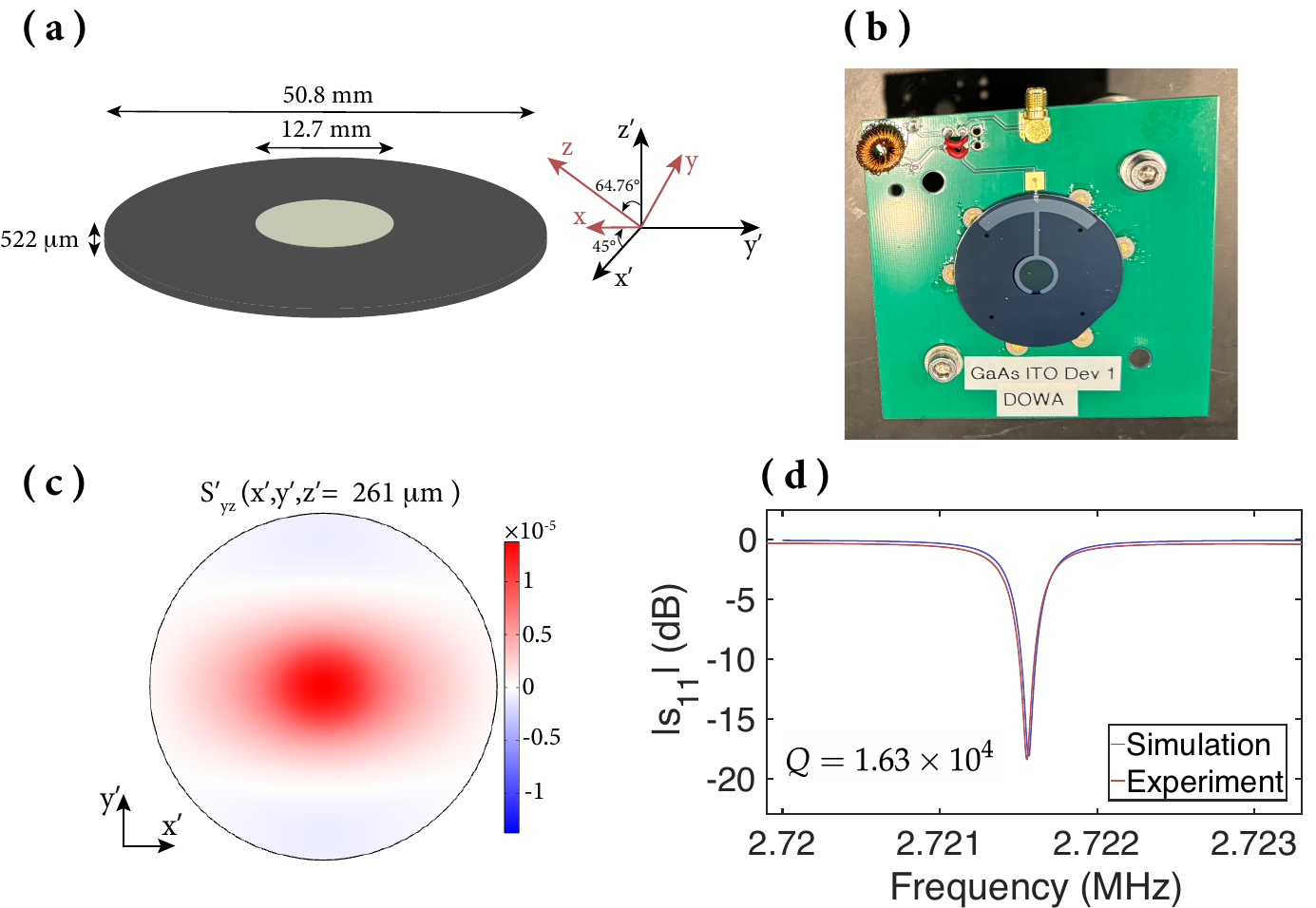}
\caption{Electromechanical characterization of the modulator. (a) Modulator dimensions and the crystal orientation are shown. (b) Fabricated modulator wirebonded and mounted on a PCB. (c) The simulated distribution of the dominant strain component amplitude $S'_{yz}(x',y',z' = 261~\mu \text{m})$ when the wafer is excited at 2.7216~MHz with 2Vpp is shown for the center of the wafer. (d) Simulated (blue) and experimental (red) of the device scattering parameter $|s_{11}|$ is shown around the fundamental shear resonance ($S'_{yz}$) for the modulator. An impedance matching transformer is mounted on the PCB and connected to the modulator.}
\label{fig:1}
\end{figure}

The modulator is constructed by depositing 140~nm of indium tin oxide (ITO) on the top and bottom surfaces of the wafer to form transparent surface electrodes with a diameter of 12.7~mm. 1~mm wide aluminum ring region on the edge of the center ITO pattern and a microstrip region that extends to the edge of the wafer is evaporated on the top and bottom surfaces of the wafer with a thickness of 200~nm. This allows the RF signal to be carried from a printed circuit board (PCB) to the center electrode region. The schematic of the fabricated modulator is depicted in Fig.~\ref{fig:1}(a). The modulator is mounted on a PCB (Fig.~\ref{fig:1}(b)), and the ends of the top and bottom aluminum microstrip regions are wirebonded to the PCB. The wafer flat is oriented along the projection of $(\hat{a}_x - \hat{a}_y)$ to the wafer surface, expressed in the crystal coordinate frame. $\hat{a}_x$ and $\hat{a}_y$ are unit vectors corresponding to directions $x$ and $y$ of the crystal, respectively, and primed notation indicates representation in the rotated coordinate frame (see the Supplement of~\cite{optically_isotropic_atalar} for more details).

\begin{figure}[ht]
\centering
\includegraphics[width=\linewidth]{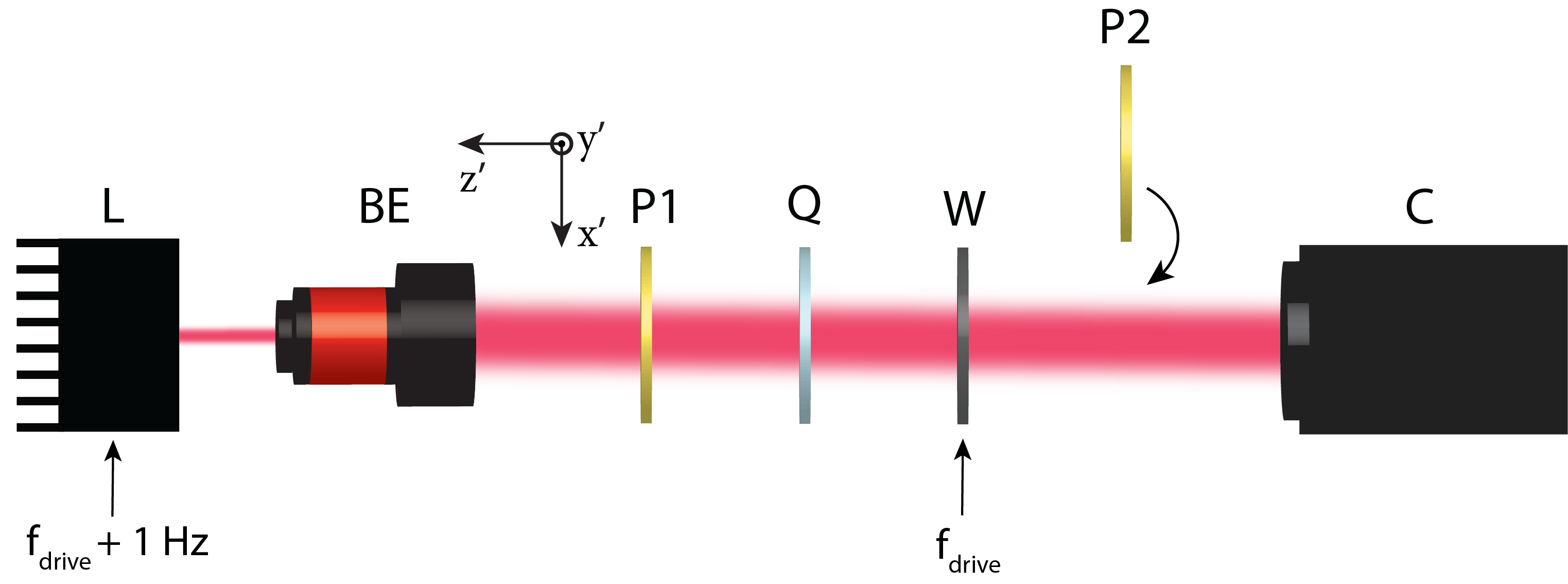}
\caption{Optical setup to characterize the modulator with a quarter-wave plate to measure the intensity modulation efficiency of the fundamental. The setup includes a laser (L) emitting light of wavelength 940~nm that is intensity modulated at $f_{drive} + 1~\text{Hz}$,
beam expander (BE) with a magnification factor of 5, two polarizers (P1) and (P2), a quarter-wave plate (Q), the modulator
(W) that is driven with an RF source of frequency $f_{drive}$, and a standard CMOS camera (C).}
\label{fig:2}
\end{figure}

\begin{figure*}[t!]
\centering
\includegraphics[width=0.78\textwidth]{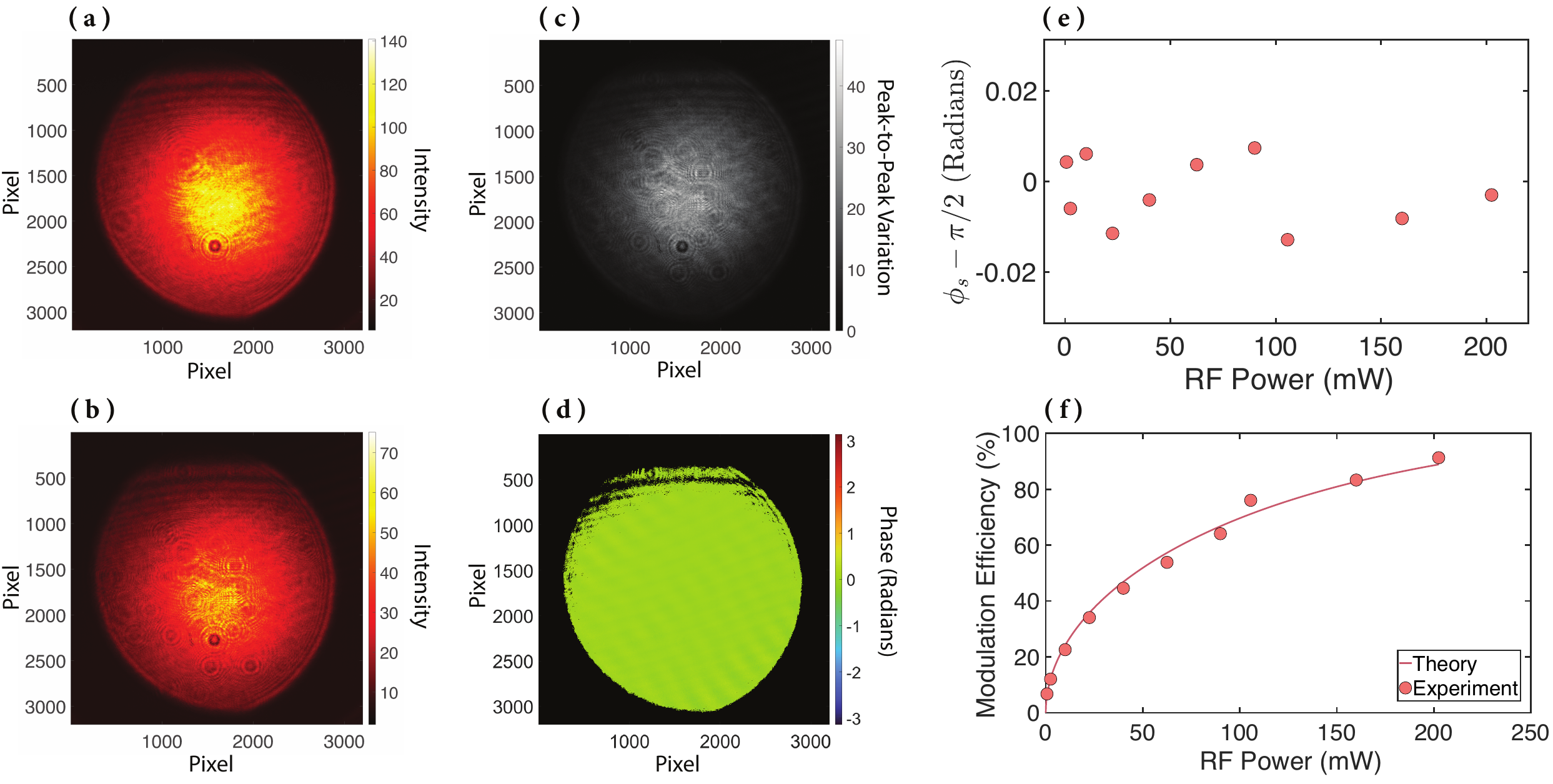}
\caption{Optical characterization results with the modulator and a quarter-wave plate for measuring the intensity modulation efficiency of the fundamental. For (a)-(d), 202.5~mW of RF power with frequency $f_{drive} = 2.720326~\text{MHz}$ is used to drive the modulator. (a) Time-averaged intensity profile of the laser beam detected by the camera is shown when (P2) is removed. (b) Time-averaged intensity profile of the laser beam detected by the camera is shown when (P2) is present. (c) The peak-to-peak variation at 1~Hz of the laser beam is shown when (P2) is present. (d) The phase of intensity modulation at 1~Hz of the laser beam is shown when (P2) is present. (e) Estimated (experimental) static birefringence of the modulator is shown for varying levels of RF excitation power. (f) Estimated (Experiment) and theoretical (Theory) modulation efficiency of the fundamental evaluated over the electrode region is shown for varying levels of RF excitation power.}
\label{fig:3}
\end{figure*}

\begin{figure}[ht]
\centering
\includegraphics[width=\linewidth]{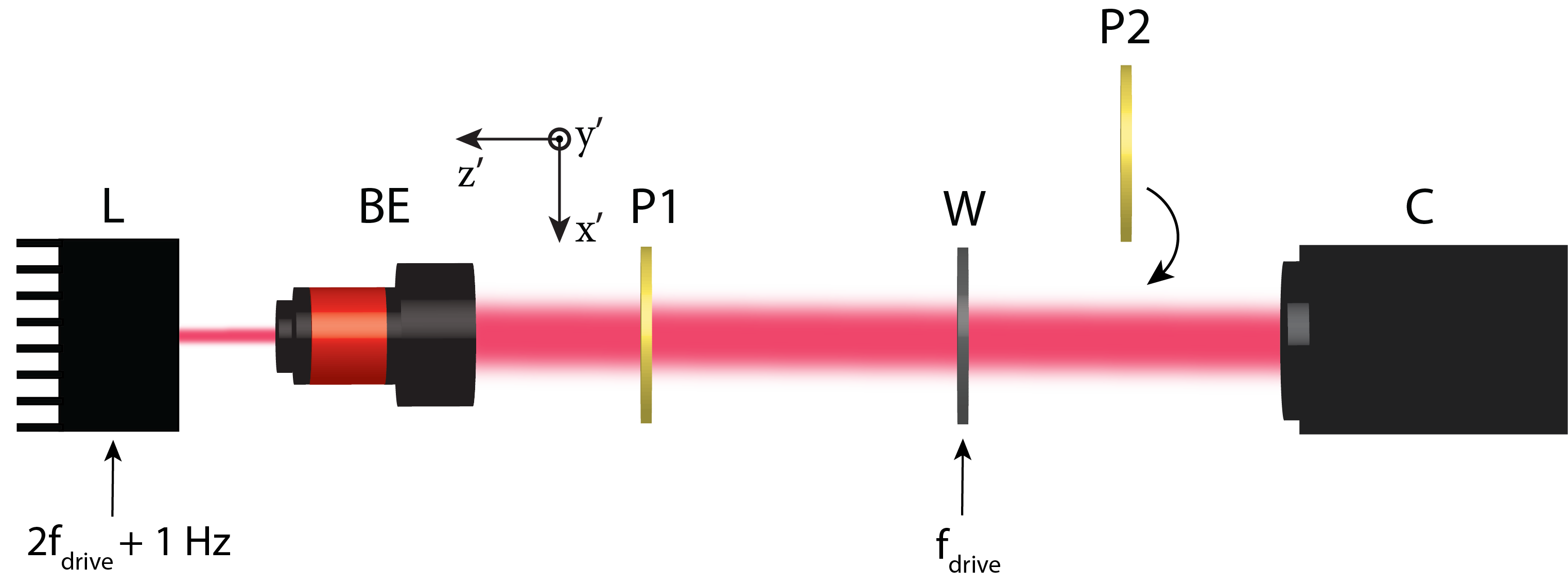}
\caption{Optical setup to characterize the modulator without a quarter-wave plate to measure the intensity modulation efficiency of the second harmonic. The setup includes a laser (L) emitting light of wavelength 940~nm that is intensity modulated at $2f_{drive} + 1~\text{Hz}$,
beam expander (BE) with a magnification factor of 5, two polarizers (P1) and (P2), the modulator
(W) that is driven with an RF source of frequency $f_{drive}$, and a standard CMOS camera (C).}
\label{fig:4}
\end{figure}

\begin{figure*}[t!]
\centering
\includegraphics[width=0.78\textwidth]{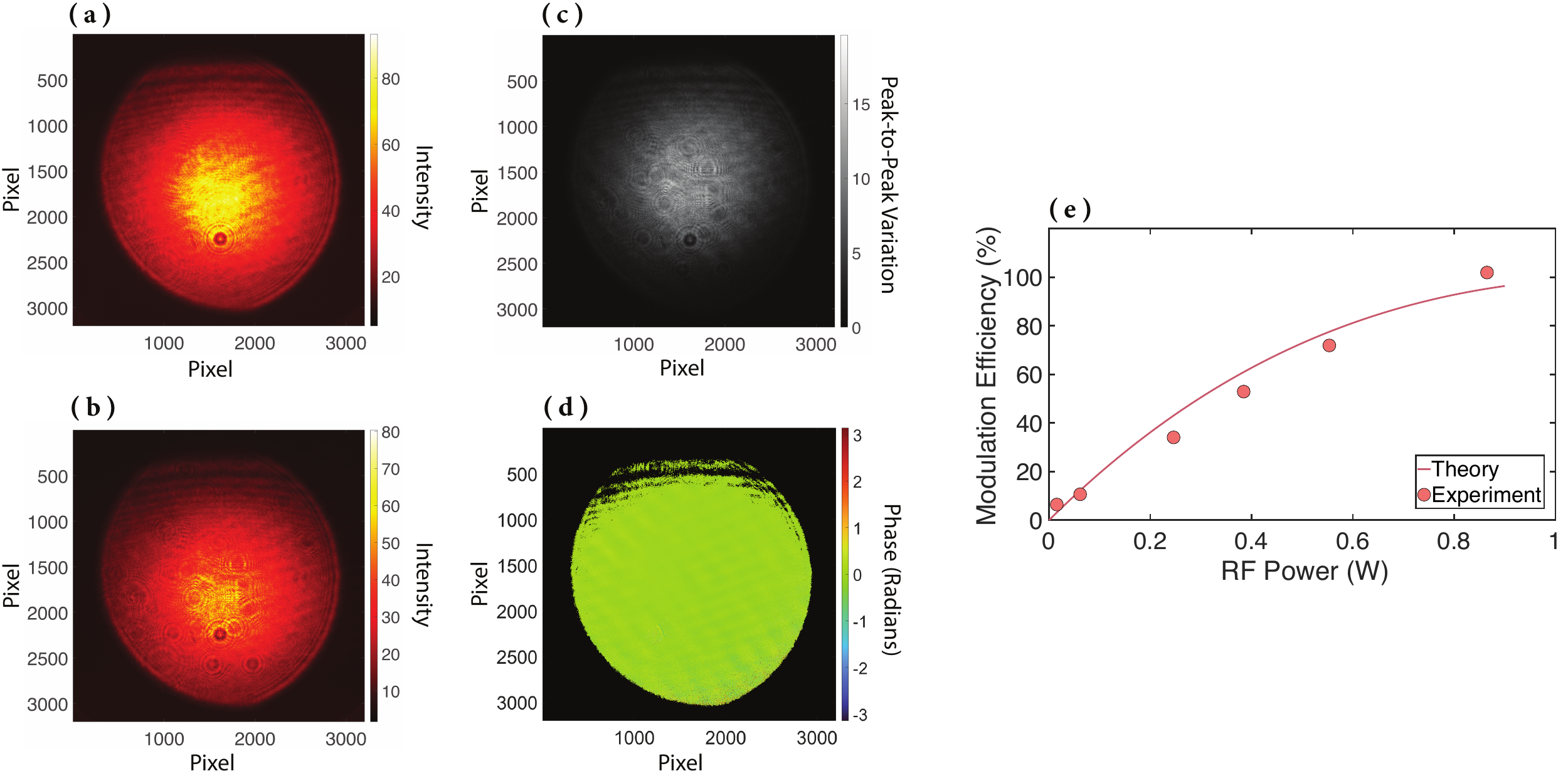}
\caption{Optical characterization results with the modulator and without a quarter-wave plate for measuring the intensity modulation efficiency of the second harmonic. For (a)-(d), 864.9~mW of RF power with frequency $f_{drive} = 2.718693~\text{MHz}$ is used to drive the modulator. (a) Time-averaged intensity profile of the laser beam detected by the camera is shown when (P2) is removed. (b) Time-averaged intensity profile of the laser beam detected by the camera is shown when (P2) is present. (c) The peak-to-peak variation at 1~Hz of the laser beam is shown when (P2) is present. (d) The phase of intensity modulation at 1~Hz of the laser beam is shown when (P2) is present. (e) Estimated (Experiment) and theoretical (Theory) modulation efficiency of the second harmonic evaluated over the electrode region is shown for varying levels of RF excitation power.}
\label{fig:5}
\end{figure*}

We first characterize the modulator electromechanically to observe the relevant fundamental shear resonanance mode ($S'_{yz}$) and estimate its $Q$. The simulated dominant strain component ($S'_{yz}$) when the wafer is driven at its fundamental shear resonance is shown in Fig.~\ref{fig:1}(c). The excited dominant strain is mostly confined to the electrode region, and exhibits high uniformity. We use a vector network analyzer (VNA) to measure the reflection scattering parameter ($s_{11}$). We extract the $Q$ for the measured mode around 2.7~MHz as $Q = 1.63 \times 10^4$ (Fig.~\ref{fig:1}(d)). Compared to our previous demonstration~\cite{optically_isotropic_atalar}, the $Q$ is higher by more than an order of magnitude and agrees very well with the simulated response (single clean mode), which we attribute to high crystal quality (different wafer supplier). To validate this, we fabricate three other modulators using three wafers from the same batch and characterize them electromechanically (see Supplementary material), observing consistent results. 

We next perform optical characterization to verify the spatial distribution of the dominant strain component $S'_{yz}(x',y',z')$ excited in the modulator and to extract the modulation efficiency. Perpendicular incidence of light to the wafer surface is used for this characterization. The excited dominant strain component $S'_{yz}$ determines the strength of the acousto-optic interaction, and consequently the degree of polarization modulation. The relation between the RF drive power ($P_{RF}$) at the fundamental shear resonance and the excited strain distribution $S'_{yz}(x',y',z')$ is as expressed in Eq.~\eqref{Eq.1}, where $f_r$ is the resonant frequency of the dominant $S'_{yz}$ strain mode, $c'_{44}$ is the rotated relevant stiffness coefficient, and $V = \pi r^2 L$ is the volume where the dominant strain is generated in the wafer ($r$ is the electrode radius and $L$ the wafer thickness). 

\begin{gather}
P_{RF} \approx \frac{4 \pi f_{r} c'_{44} \int_V S'^2_{yz}(x',y',z')dV}{Q}.  \label{Eq.1}
\end{gather}

The relation between the root mean square (rms) polarization modulation amplitude (${\phi_{D}}_{rms}$) in radians and the excited strain distribution $S'_{yz}(x',y',z')$ is expressed in Eq.~\eqref{Eq.2}, where $\lambda$ is the free-space wavelength of light, $n$ is the refractive index of GaAs, $p'_{14}$ and $p'_{24}$ are the rotated relevant photoelastic coefficients, and $A = \pi r^2$ is the active area of the modulator. 

\begin{gather}
{\phi_{D}}_{rms} \approx \int_{0}^L \frac{2 \pi n^3}{\lambda}\Big(p'_{14} - p'_{24}\Big)\sqrt{\frac{\int_A S'^2_{yz}(x',y',z')dA}{\pi r^2}}dz'. \label{Eq.2}
\end{gather}

When the modulator is placed between polarizers (with transmission axis $\frac{\hat{a}'_x + \hat{a}'_y}{\sqrt{2}}$) to excite the two eigenmodes equally (along $x'$ and $y'$ directions) of the crystal, polarization modulation is converted to intensity modulation, where the relative amplitudes of the excited harmonic terms are scaled by the Bessel functions of the first kind. Due to the lack of birefringence of the modulator, the odd or the even harmonic intensity modulation components could be selected by the presence or absence of a quarter-wave plate, respectively (see the Supplement of \cite{longitudinal_nat_paper} Eq.~S(38) for the full expression). We first measure the fundamental intensity modulation term at approximately 2.7~MHz using a quarter-wave plate with its fast axis oriented along $\hat{a}'_x$. The setup depicted in Fig.~\ref{fig:2} is used, where two separate captures are made with the second polarizer placed in and out of the laser beam path. When the second polarizer (P2) is in the laser beam path, the rms intensity detected by the camera is expressed as Eq.~\eqref{Eq.3}, where $I_{0rms}$ is the rms intensity of the unpolarized light emitted by the laser, $J_1$ the first Bessel function of the first kind, $t$ is time, and HOH stands for the higher order odd harmonics. When P2 is removed, $I_{0rms}/2$ could be measured, allowing calibration and extraction of $\phi_{Drms}$. 

\begin{gather}
I_{rms1}(t) \approx \frac{{I_0}_{rms}}{2}\Big(\frac{1}{2} - J_1\big({\phi_{D}}_{rms}\big)\text{cos}(2 \pi t f_{drive}) \Big) + \text{HOH}. \label{Eq.3}
\end{gather}

The modulator is driven with RF frequency $f_{drive}$ that is selected to match the fundamental shear resonance frequency $f_r$ of the modulator. This generates the dominant strain component $S'_{yz}$ in the wafer. Due to heating of the wafer when the device is driven, $f_r$ varies and $f_{drive}$ is adjusted to track it. A negative feedback loop with a directional coupler and an oscilloscope is used for frequency stabilization of the modulator (similar to the method used in ~\cite{optically_isotropic_atalar}). A laser diode emitting light of wavelength 940~nm (where GaAs is transparent) is used as the light source. The light source is intensity modulated at $f_{drive} + 1~\text{Hz}$ to enable heterodyne detection for the 1~Hz beat tone with the camera to extract the modulation strength, where 80 frames are captured with a frame rate of 4~Hz for the two measurements. The captures are made after thermal equilibrium is reached and $f_{drive}$ matches $f_r$ using the frequency stabilization method.

The optical characterization results with the quarter-wave plate are shown in Fig.~\ref{fig:3} for varying levels of RF drive power. We observe spatial phase uniformity across the laser beam (as expected, Fig.~\ref{fig:3}(d)), and modulation efficiency reaching 100\%, translating to rms polarization modulation amplitude $\phi_{Drms}$ of 1.28 radians with only 200~mW of RF drive power (Fig.~\ref{fig:3}(f)). The modulation efficiency is calculated by extracting $J_1({\phi_{Drms}})$ and normalizing to the maximum of $J_1$. The experimentally calculated modulation efficiency agrees well with the linear photoelastic interaction model (theory). We also estimate the static birefringence ($\phi_s$), which is expected to be equal to $\pi/2$ radians due to the presence of the quarter-wave plate (Fig.~\ref{fig:3}(e)). The deviation of $\phi_s$ for different drive powers is less than 0.02~radians from the baseline value of 0 radians, which experimentally validates the lack of birefringence for the modulator. 

We next perform optical characterization without the quarter-wave plate, allowing us to measure the second harmonic intensity modulation term (around 5.4~MHz) resulting from the polarization modulation coupled with polarizers. The setup depicted in Fig.~\ref{fig:4} is used, where two separate captures are made with the second polarizer placed in and out of the laser beam path. When the second polarizer (P2) is in the laser beam path, the rms intensity detected by the camera is expressed as Eq.~\eqref{Eq.4}, where $J_0$ and $J_2$ are the zeroth and second Bessel functions of the first kind, respectively, and HOH stands for the higher order even harmonics. When P2 is removed, $I_{0rms}/2$ could be measured, allowing calibration and extraction of $\phi_{Drms}$. 

\begin{gather}
I_{rms2}(t) \approx \frac{{I_0}_{rms}}{2}\bigg(\frac{1}{2} + \frac{1}{2}J_0\big({\phi_{D}}_{rms}\big) \nonumber \\  - J_2\big({\phi_{D}}_{rms}\big)\text{cos}(4 \pi t f_{drive}) \bigg) + \text{HOH}. \label{Eq.4}
\end{gather}

The light source is intensity modulated at $2f_{drive} + 1~\text{Hz}$ to enable heterodyne detection for the 1~Hz beat tone with the camera to extract the modulation strength, where 80 frames are captured with a frame rate of 4~Hz for the two measurements. The optical characterization results without the quarter-wave plate are shown in Fig.~\ref{fig:5} for varying levels of RF drive power. We again observe phase uniformity across the laser beam (as expected, Fig.~\ref{fig:5}(d)), and modulation efficiency reaching 100\%, translating to rms polarization modulation amplitude $\phi_{Drms}$ reaching 2.7 radians (corresponding to rms $S'_{yz}$ of $3.5 \times 10^{-4}$) with less than 1~W of RF drive power. The modulation efficiency is calculated by extracting $J_2({\phi_{Drms}})$ and normalizing to the maximum of $J_2$. The experimentally calculated modulation efficiency agrees relatively well with the linear photoelastic interaction model (theory). Using the optical characterization setup, we also measure the optical insertion loss for perpendicular incidence of light as 1.2~dB for the modulator. 

Photoelastic modulators are widely used optical devices functioning as free-space resonant polarization modulators for various applications. The transverse acousto-optic interaction mechanism underlying the operation of these devices leads to a fundamental trade-off between the input aperture and the modulation frequency, with commercially available modulators operating near 50~kHz for centimeter-square apertures. In this work, using a collinear acousto-optic interaction mechanism, we demonstrated a photoelastic modulator operating at 2.7~MHz with a centimeter square aperture. This marks a greater than a factor of 20 improvement compared to commercially available devices. We demonstrated substantial polarization modulation, close to $\pi$ radians in amplitude, with sub-watt drive power which was verified by demonstrating nearly 100\% intensity modulation efficiency at the fundamental and second harmonic frequencies of the fundamental shear resonance of the modulator. The high operating frequency, large aperture, sub-millimeter thickness, high modulation efficiency, and low drive power make the presented modulator promising for a plethora of applications.


\section*{Funding.}
Stanford SystemX Alliance.

\section*{Acknowledgment.}
The authors thank Prof. Amir H. Safavi-Naeini and Prof. Butrus (Pierre) T. Khuri-Yakub for providing lab space to conduct the experiments.

\section*{Disclosures.}
The authors declare no conflicts of interest

\section*{Data availability.}
Data underlying the results presented in this paper are
not publicly available at this time but may be obtained from the authors upon
reasonable request.

\bibliographystyle{unsrt}
\bibliography{references}


\onecolumn
\newpage


\section*{\textbf{\fontsize{16}{19.2}\selectfont Supplementary material}}

\bigskip \bigskip 

\renewcommand\thefigure{S\arabic{figure}}
\setcounter{figure}{0} 

\setcounter{section}{0}

\section*{Electromechanical characterization of the fabricated modulators}
The modulators are produced from wafers within the same batch, each having a diameter of $50.8 \pm 0.3$~mm and a thickness of $0.5 \pm 0.02$~mm. These wafers are undoped, double-side polished, and have a crystal orientation of $(332) \pm 1^{\circ}$ in Miller indices notation. For two of the modulators, 100 nm thick aluminum electrodes are deposited on both the top and bottom surfaces. For the other two modulators, 140 nm thick ITO is deposited on the top and bottom surfaces.

\begin{figure*}[h!tbp]
\centering
\includegraphics[width=1\textwidth]{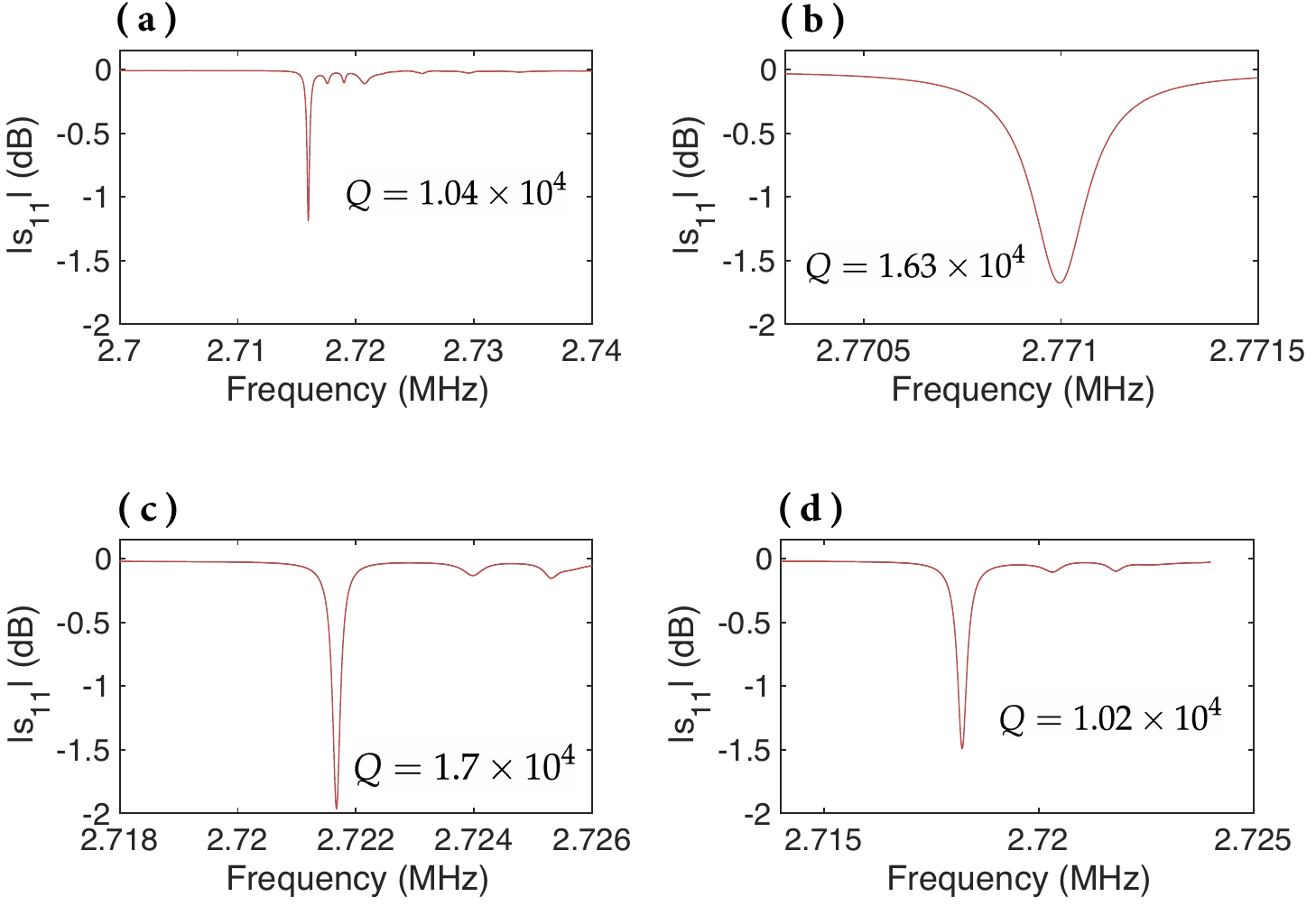}
\caption{Electromechanical characterization of the fabricated modulators. The device scattering parameter $|s_{11}|$ is measured around the fundamental shear resonance ($S'_{yz}$) for the modulators. (a) First modulator fabricated by depositing 100~nm thick aluminum electrodes. (b) Second modulator fabricated by depositing 100~nm thick aluminum electrodes. (c) Third modulator fabricated by depositing 140~nm thick ITO electrodes. (d) Fourth modulator fabricated by depositing 140~nm thick ITO electrodes.}
\label{fig:s1}
\end{figure*}

A total of four modulators are fabricated and electromechanically characterized by measuring the $s_{11}$ scattering parameter near the fundamental shear resonance frequency of the wafer using a VNA. The measured responses are shown in Fig.~\ref{fig:s1}. The resonance frequency location and response of the four modulators are generally consistent, with some variations due to uncertainties in wafer parameters. The modulator characterized optically in the main manuscript is the one depicted in (c) (without an impedance matching transformer). The $Q$ is slightly higher without the impedance matching transformer, as expected.

\end{document}